\documentclass[useAMS]{mn2e}
\usepackage{graphicx}

\def\simgt{\ {\raise-.5ex\hbox{$\buildrel>\over\sim$}}\ }

\def\cd{d$^{-1}$\,}
\def\kms{kms$^{-1}$\,}

\begin{document}

\title[Variability of the central star of NGC 6826]
{{\it Kepler} photometry and optical spectroscopy of the ZZ Lep central star 
of the Planetary Nebula NGC 6826: rotational and wind variability}
\author[G. Handler et al.]
{G. Handler,$^{1}$ R. K. Prinja,$^{2}$ M. A. Urbaneja,$^{3}$ 
V. Antoci,$^{4, 5}$ J. D. Twicken,$^{6}$ \and T. Barclay\,$^{7}$
\and \\
$^1$ Copernicus Astronomical Center, Bartycka 18, 00-716 Warsaw, Poland (gerald@camk.edu.pl)\\
$^{2}$ Department of Physics \& Astronomy, University College London, 
Gower Street, London WC1E 6BT\\
$^{3}$ Institute for Astronomy, University of Hawaii, 2680 Woodlawn 
Drive, Honolulu, HI 96822, USA\\
$^{4}$ Stellar Astrophysics Centre (SAC), Department of Physics and 
Astronomy, Aarhus University, Ny Munkegade 120,\\ DK-8000 Aarhus C, 
Denmark\\
$^{5}$ Department of Physics and Astronomy, University of British 
Columbia, Vancouver, BC V6T1Z1, Canada\\
$^{6}$ SETI Institute/NASA Ames Research Center, Moffett Field, CA 
94035, USA\\
$^{7}$ Bay Area Environmental Research Inst./NASA Ames Research Center, 
Moffett Field, CA 94035, USA}

\date{Accepted 2012 July 17.
 Received 2012 August 13;
 in original form 2012 September 10}
\maketitle

\begin{abstract}
We present three years of long-cadence and over one year of 
short-cadence photometry of the central star of the Planetary 
Nebula NGC 6826 obtained with the {\it Kepler} spacecraft, and 
temporally coinciding optical spectroscopy. The light curves are 
dominated by incoherent variability on time scales of several hours, but 
contain a lower-amplitude periodicity of 1.23799~d. The temporal 
amplitude and shape changes of this signal are best explicable with a 
rotational modulation, and are not consistent with a binary 
interpretation. We argue that we do not observe stellar pulsations 
within the limitations of our data, and show that a binary central star 
with an orbital period less than seven days could only have escaped 
our detection in the case of low orbital inclination. Combining the 
photometric and spectroscopic evidence, we reason that the hourly 
variations are due to a variable stellar wind, and are global in nature. 
The physical cause of the wind variability of NGC 6826 and other ZZ 
Leporis stars is likely related to the mechanism responsible for wind 
variations in massive hot stars.
\end{abstract}

\begin{keywords}
stars: variables: other -- stars: early-type -- stars: mass loss --
stars: rotation -- stars: winds, outflows -- planetary nebulae: 
individual: NGC 6826
\end{keywords}

\section{Introduction}

The ZZ Leporis stars (Handler 2003) are a group of fourteen variable 
Central Stars of Planetary Nebulae (CSPN), named after their prototype, 
the central star of IC~418. The cause of their variability remains to be 
fully understood. Photometric variations are present on two time scales, 
of the order of days and of the order of hours. The most plausible 
mechanisms for the variability are pulsation or variations in the 
stellar mass loss rate - or both. Binary-induced light variations can in 
most cases be excluded due to the observed light curve shapes, the lack 
of a dominant periodicity, and the lack of corresponding radial velocity 
changes.

The reason why pulsations may be expected to be present in ZZ Lep stars 
is evident from their positions in the HR diagram. They are located at 
the intersection of the post-Asymptotic Giant Branch evolutionary track 
with the instability strip of the $\beta$~Cephei pulsators. Gautschy 
(1993) showed that stellar models reminiscent of those stars have 
pulsationally unstable eigenmodes. Zalewski (1993) confirmed these 
results and showed that nonlinear models in the temperature range of the 
ZZ Lep stars undergo quite complicated light and radial velocity 
variability on time scales of several hours.

Wind variability, on the other hand, was proposed as a mechanism by 
M\'endez, Verga \& Kriner (1983). UV spectroscopic studies (e.g., Prinja 
et al.\ 2012) support this idea, because all ZZ Lep stars examined 
indeed showed clear evidence for variable winds. It is believed that the 
temporal behaviour of the winds of those CSPN is similar to that of 
massive OB stars (e.g., Kaper et al.\ 1997), is related to the stellar 
rotational period, and may be rooted close or at the stellar surface 
(e.g., Fullerton et al.\ 1997). Further strong support for this 
interpretation was recently presented by Prinja, Massa \& Cantiello 
(2012), who also argued that the wind variations of the central star of 
NGC~6543 could be causally connected to sub-surface convection. The 
radiation pressure driven winds of hot stars generally appear to exhibit 
extensive variability in their stellar winds, whether on the main 
sequence, as supergiants, Wolf-Rayet stars or CSPN.

Whatever the cause of variability of the ZZ Lep stars, it will have 
implications for our astrophysical knowledge of the evolution of CSPN. 
They cross the HR diagram on time scales of a few thousand years under 
the influence of mass loss. If ZZ Lep stars pulsate, measurements of 
oscillation period changes could closely constrain their evolutionary 
speed. Model calculations (e.g., Bl\"ocker 1995) imply that this speed 
is highly mass dependent and therefore CSPN masses could be accurately 
determined. If the complicated light variations of ZZ Lep stars were due 
to multimode pulsations, they might even be accessible to 
asteroseismology. Constraints on CSPN mass loss behaviour, on the other 
hand, are also an asset for understanding their evolution, spectral 
characteristics and even for the shaping of the surrounding nebulae 
(e.g., Huarte-Espinosa et al. 2011).

One of the reasons why the variability mechanism of the ZZ Lep stars is 
so hard to pinpoint may be their complicated behaviour in combination 
with a lack of supporting data. Ground-based observations suffer from 
gaps due to daylight interruptions (even in case of multisite 
campaigns), whereas time on rather large telescopes or UV spacecraft 
required for spectroscopic monitoring is very limited. In addition, ZZ 
Lep stars are rather faint and surrounded by bright nebulae that 
complicate measurements and their interpretation. On the other hand, 
their variations occur on shorter time scales than those of massive OB 
stars, which are of the order of one day, which can be an observational 
advantage.

One ZZ Lep star is located within the field of view of the {\it Kepler} 
mission (Koch et al.\ 2010). HD 186924 = KIC 12071221 ($V=10.4$) is the 
central star of NGC 6826\footnote{In the remainder of this paper, we 
will use the name of the Planetary Nebula also as the designation of the 
central star.}. Its photometric variability was first reported by Bond 
\& Ciardullo (1990) and confirmed by Handler (1998) and Michalska 
(2000). The light range of the variations was several hundredths of a 
magnitude, and the time scale several hours, with no clear evidence for 
a periodicity. UV spectroscopy (Prinja et al.\ 2012) revealed the 
presence of recurring Discrete Absorption Components (DACs) in the 
absorption troughs of wind-sensitive lines, clear evidence for variable 
wind structure.

Jevti\'{c} et al.\ (2012) examined the first $\sim$300\,d of {\it Kepler} 
photometry of NGC 6826 by means of nonlinear time series analysis and 
phase-space reconstruction. They noted that the variability is coherent 
on time scales of up to about 1.5\,hr, but becomes consistent with noise 
on time scales exceeding 10\,hr. In the present work, we analyse all 
presently available {\it Kepler} time-series photometry with more traditional 
techniques. Combining the photometric results with constraints from
optical time-series spectroscopy, we discuss interpretations of the
central star variability.

\section{Observations}

\subsection{Photometry}

{\it Kepler} photometry is provided in two modes, short cadence (SC) and long 
cadence (LC). In SC mode (Gilliland et al.\ 2010a), a data point is 
obtained every 58.8 seconds, whereas LC mode (Jenkins et al.\ 2010b) 
yields a data point every 29.4 minutes. The data are arranged in 
quarters of three months each, with the exception of the first two. Q0 
consists of 10 days of commissioning data and Q1 of one month of science 
data. In between the quarters there are gaps of about one day for data 
download and spacecraft rotation.

NGC 6826 was and still is observed in LC mode from Q1 (starting on 13 
May 2009) onwards. After examination of the first LC data, SC 
measurements were requested within the Kepler Asteroseismic Science 
Consortium (KASC, Gilliland et al. 2010b), which were obtained from Q7 - 
Q11. {\it Kepler} photometry is available as ``corrected" and 
``uncorrected" data. In this sense, ``uncorrected" means that the data 
have undergone standard reductions such as bias correction, 
flat-fielding and correction for the smear induced by the readout, 
whereas it has been tried to remove instrumental systematics in the 
``corrected" data. We started from the ``uncorrected" data because 
a comparison showed that these will result in a lower noise level. Both 
the LC and SC light curves were first converted to magnitudes, 
visually inspected, and in case of the presence of simultaneous LC and 
SC data compared. Long-term drifts were assumed to be instrumental (for 
instance, arising from differential velocity aberration) and were 
filtered out with polynomials up to third order to account for the 
instrumental drifts properly whilst avoiding to ``overfit'' the data in 
one-month data chunks. Hence, our analysis will not be sensitive to 
periodicities longer than about a week. Some very few outliers and bad 
sections of data, e.g.\ after pointing losses of the spacecraft whose 
times are documented were removed by hand. The final data set on which 
this paper is based comprises the Q1 - Q12 LC data, which span 1051 days 
with a duty cycle of 93.3\%, and the SC data containing 468 days of 
observation with a 92.3\% duty cycle. We estimate a precision per 
data point of 0.17 mmag for the SC data; the apparent precision of the 
LC data however only is 0.43 mmag per point, implying that most of the 
``noise" in these measurements is actually signal.

One difficulty with CSPN photometry is the presence of the nebula (that 
is not expected to vary on the time scales investigated here). Its 
influence can be suppressed by choosing filters at low nebular continuum 
flux, and avoiding strong nebular emission lines. Another method is to 
mathematically subtract the nebular contribution or image subtraction. 
All these are not suitable for the {\it Kepler} data. We use the normal Simple 
Aperture Photometry (SAP, Jenkins et al.\ 2010a) data. Those contain 
some nebular flux but are more robust against artificial amplitude 
changes coming from e.g.\ (quarterly) pointing changes of the spacecraft 
than single-pixel data. As a consequence, the amplitudes of the light 
variations reported here are smaller than they intrinsically are. A 
comparison between SAP and pixel data implies an amplitude reduction to 
less than one quarter of the intrinsic value. A subset of the SC data is 
displayed in Fig.\ 1.

\begin{figure*}
\includegraphics[angle=270,width=177mm,viewport=0 3 295 740]{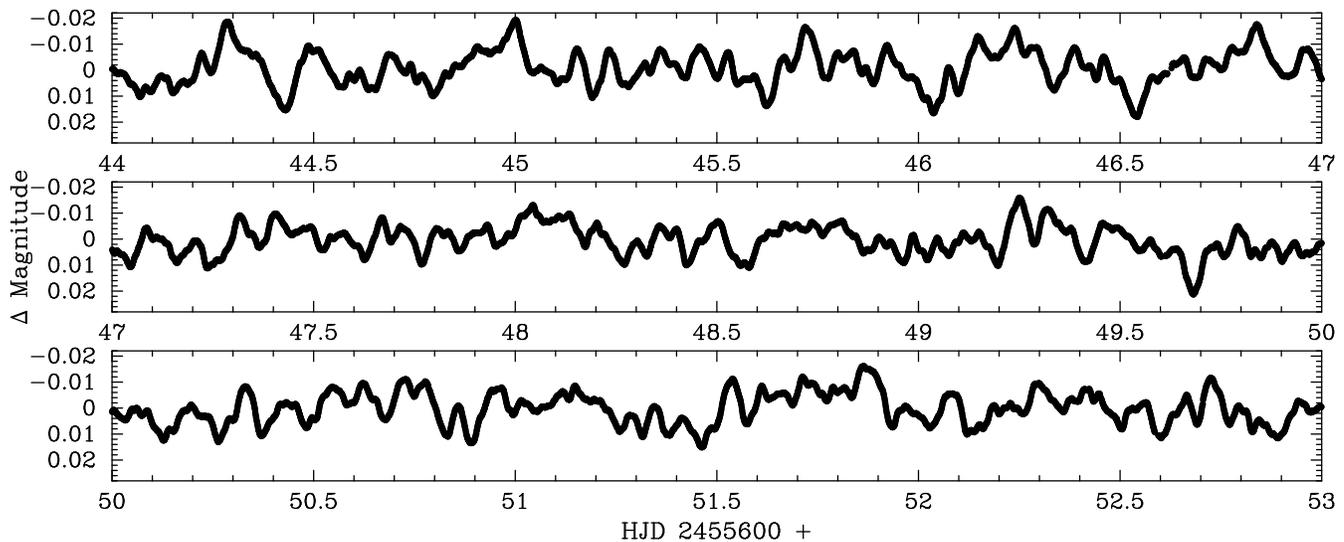}
\caption[]{A nine-day portion of the {\it Kepler} photometry of NGC 6826.}
\end{figure*}

\subsection{Spectroscopy}

We obtained high-resolution spectroscopy co-temporal with a section of 
the {\it Kepler} photometry. The first data set stems from the 3.6-m 
Canada France Hawaii Telescope (CFHT) and the ESPaDOnS spectrograph 
(Donati 2003). Observations were taken on three consecutive nights from 
30/31 May to 1/2 June, 2010 (UT), totalling 14 spectra with 
individual exposure times of 1800\,s, a wavelength coverage of 3700 to 
10480 \AA{} and a spectral resolution of 68000 in ``star + sky" 
mode. On the 2.6-m Nordic Optical Telescope (NOT), the FIES (Frandsen 
\& Lindberg 1998) spectrograph was used on the nights of 6/7 to 
7/8 June, 2010 (UT).  Thirteen spectra were gathered with 
individual exposure times of 1800\,s, in a wavelength range of 3640 to 
7360\,\AA{} with $R=46000$.

The optical spectrum of NGC 6826 is characterised by variable lines 
including He{\sc II} 4542 \AA\, and He{\sc II} 5411 \AA\, in absorption. 
N{\sc V} 4605 \AA\, exhibits a weak P Cygni profile, and He{\sc II} 4686 
\AA\, as well as C{\sc IV} 5801\AA\, and 5812 \AA\, are in strong 
(stellar) emission. Further visible are the Balmer absorption wings (the 
line cores are contaminated by nebular emission), and N{\sc IV} 
6381\AA\, is a rare, but very weak, photospheric line. Figure 2 shows 
examples of the hourly variability present in He{\sc II} 4686 \AA\, and 
He{\sc II} 5411 \AA\, in the CFHT and NOT spectra (see Sect. 3.2).

\begin{figure}
\includegraphics[width=84.3mm,viewport=17 20 417 740]{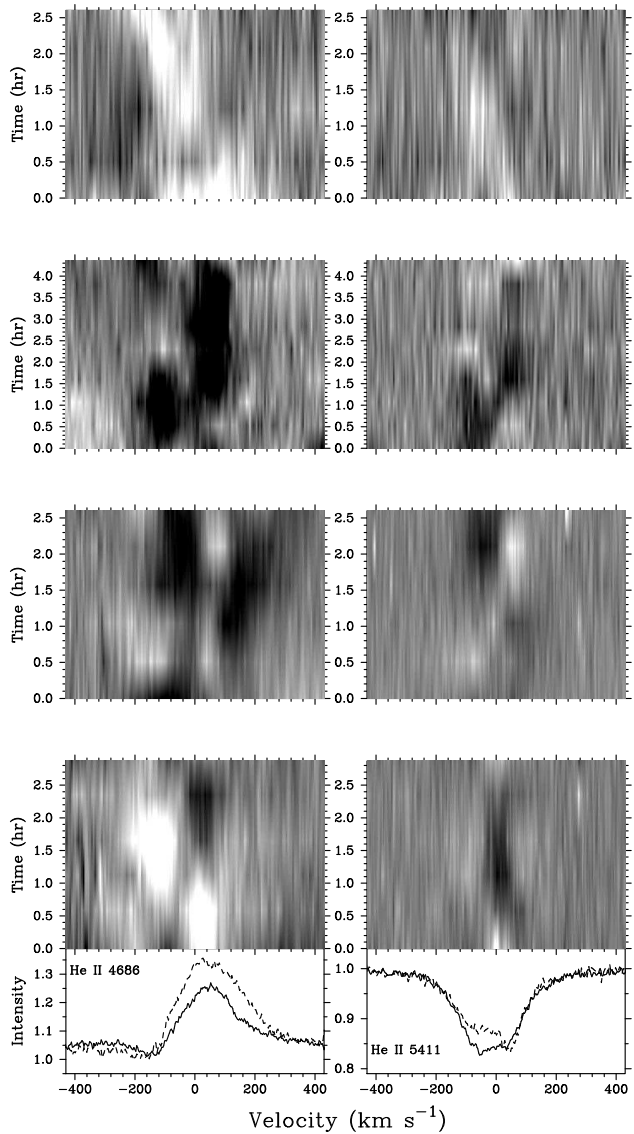}
\caption[]{Dynamic spectra of the He{\sc II} 4686 \AA\, and He{\sc II} 
5411 \AA\, lines. The individual spectra have been ratioed by the mean 
spectrum for each telescope. The top two panels are NOT data, whereas 
the bottom two represent the CFHT spectra. The panel at the very 
bottom shows examples of individual spectra from the CFHT data to 
illustrate the maximum extent of the line intensity changes. There is 
some correspondence between the variability patterns seen in these two 
He{\sc II} lines that is also seen in some other spectral features. He{\sc 
II} 4686 is variable to at least $\pm 200$\,km/s.}
\end{figure}

\section{Analysis}

\subsection{Photometric frequencies}

The light variations of NGC 6826 are quite complicated, predominantly
occur on time scales of a few hours and show no obvious periodicity
(Fig.\ 1). We analysed the whole SC data set using the {\sc Period04}
software (Lenz \& Breger 2005). This package applies single-frequency
power spectrum analysis and simultaneous multi-frequency sine-wave
fitting. It also includes advanced options such as the calculation of
optimal light-curve fits for multiperiodic signals including harmonic,
combination, and equally spaced frequencies.

\begin{figure}
\includegraphics[width=84.3mm,viewport=0 0 340 216]{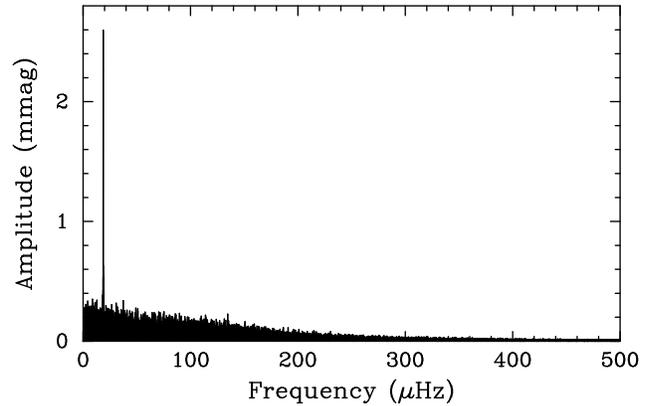}
\caption[]{Amplitude spectra of the short-cadence {\it Kepler} photometry of NGC 
6826.}
\end{figure}

The amplitude spectrum of the SC data is shown in Fig.\ 3. It is 
dominated by a single frequency, surrounded by apparent noise dropping 
continuously in amplitude towards higher frequencies. No coherent signal 
is present between 20 $\mu$Hz up to the Nyquist frequency of 8500 
$\mu$Hz.

\begin{figure}
\includegraphics[width=84.3mm,viewport=0 0 340 304]{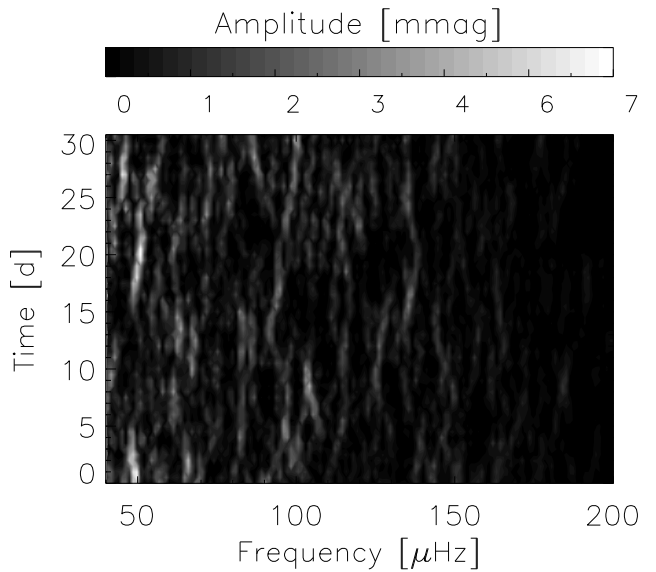}
\caption[]{Time-Fourier spectrum of one month of short-cadence {\it Kepler} 
photometry (Q9.1) of NGC 6826, with a sliding window size of 5\,d. To 
accentuate the variations on short time scales, the dominating 
low-frequency signal is not shown.}
\end{figure}

However, the predominant light variations of NGC 6826 seem to occur on 
this time scale (Fig.\ 1). Therefore, we computed Time-Fourier spectra 
of all the SC data. One example of those is shown in Fig.\ 4, 
where we chose a five-day sliding window and a time step of 0.5 days. 
Again, there is no evidence for coherent or recurrent signals. If there 
is power in the Time-Fourier spectrum, it is short-lived, as 
experimenting with different temporal window sizes prove. In Fig.\ 4 it 
is also visible that the amplitudes and frequencies of the short-term 
variations are not stable in time, but are variable as well. As 
there is no more astrophysical information to be extracted from the 
Time-Fourier analysis, we decided to continue our investigations with 
the simplest possible methods.

As an additional test, we divided the data set into two halves and 
computed their amplitude spectra. Their appearance is consistent with 
that expected from noise: their background levels are similar, but 
compared to the full data set by a factor of about $\sqrt{2}$ higher. 
This result is fully consistent with the earlier study by Jevti\'{c} et 
al.\ (2012), who showed that the variations are coherent over short time 
scales only.

\begin{figure}
\includegraphics[width=84.3mm,viewport=0 0 343 332]{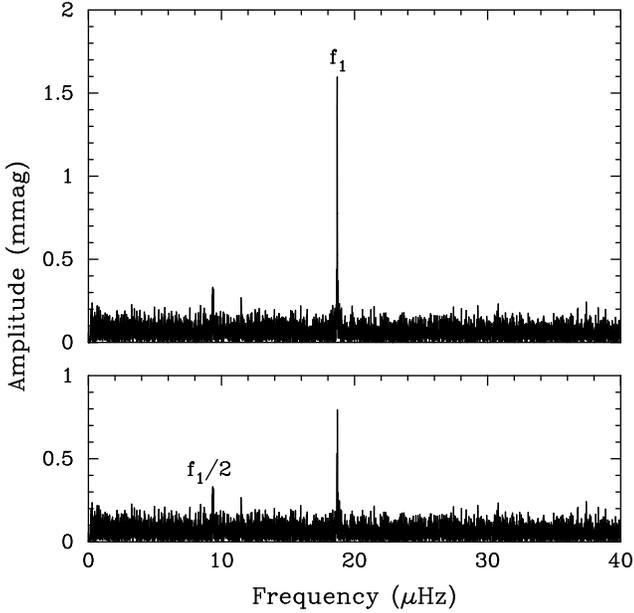}
\caption[]{Amplitude spectra of the $\sim 3$-year LC {\it Kepler} photometry 
in the low-frequency domain. The dominant signal $f_1$ does not cleanly 
prewhiten. In the lower panel, the subharmonic of $f_1$ also stands out.}
\end{figure}

Turning to the apparently coherent signal, we can make use of the LC 
data and their longer time base. The amplitude spectrum of these 
measurements is examined in Fig.\ 5 (upper panel). As in Fig.\ 3, there 
is a dominant signal, but its amplitude is only about 60\% of that in 
the SC data.  Prewhitening this signal (lower panel of Fig.\ 5) does not 
remove all the power around this frequency; several peaks in a narrow 
interval ($\sim 0.06\,\mu$Hz) remain. In addition, a subharmonic of this 
signal is detected. Investigating the subharmonic with the same 
methodology shows a similar picture: it is not reproducible with a 
single coherent frequency, and residual peaks in an $\sim 0.09\,\mu$Hz 
interval remain after prewhitening.

To assess the significance of the detection of the two signals, we 
adopted the widely used and reliable criterion by Breger et al.\ (1993). 
According to this criterion, any peak that exceeds a signal-to-noise 
(S/N) ratio of four in the amplitude spectrum is statistically 
significant. To compute the noise level in the presence of red noise and 
time-variable signals, we integrated over the amplitude spectrum between 
$0-8 \mu$Hz and $20-28 \mu$Hz and interpolated the results to the 
frequencies of interest. The dominant peak has $S/N$ of 22.5, and its 
subharmonic of 4.4. The next tallest peak in Fig.\ 5 has $S/N=3.6$ and 
is therefore not significant.

These findings were traced to a change of the light curve amplitude and 
shape over monthly time scales. Phase diagrams of 2-month 
stretches of data, folded among the subharmonic frequency are displayed 
in Fig.\ 6. Averaging over such time intervals was found most suitable 
to suppress the incoherent short-term variations whilst sampling the 
temporal change of the periodic component of the light curve well
and not affecting any scientific conclusions drawn. The light curve 
evolved from a double-wave shape to a single-wave variation, then almost 
dropped to zero amplitude and reached almost a 0.01 mag light range 
before dropping to an asymmetric double-wave variation again. The phases 
of light maximum and minimum varied only slightly over the three years 
of {\it Kepler} observations. 

We adopted the subharmonic frequency $9.3491\pm0.0001\,\mu$Hz,
corresponding to a period of $1.23799\pm0.00001$\,d, as the fundamental
time scale of the regular variability of NGC 6826. The reason for
this choice is that in almost all variable stars in whose amplitude
spectra a subharmonic is detected, it is the subharmonic frequency that is
related to the physical cause of the variability. Prominent examples are
eclipsing binaries or ellipsoidal and rotational variables. The only
exception from this rule is the period doubling phenomenon, a rare and
complex resonance effect in some pulsating stars (e.g., see Szab\'o et
al.\ 2010). Aside from that, adopting the tallest peak in the amplitude
spectrum as the physical variability time scale would leave the
subharmonic unexplained.

\begin{figure}
\includegraphics[width=84.3mm,viewport=0 0 301 727]{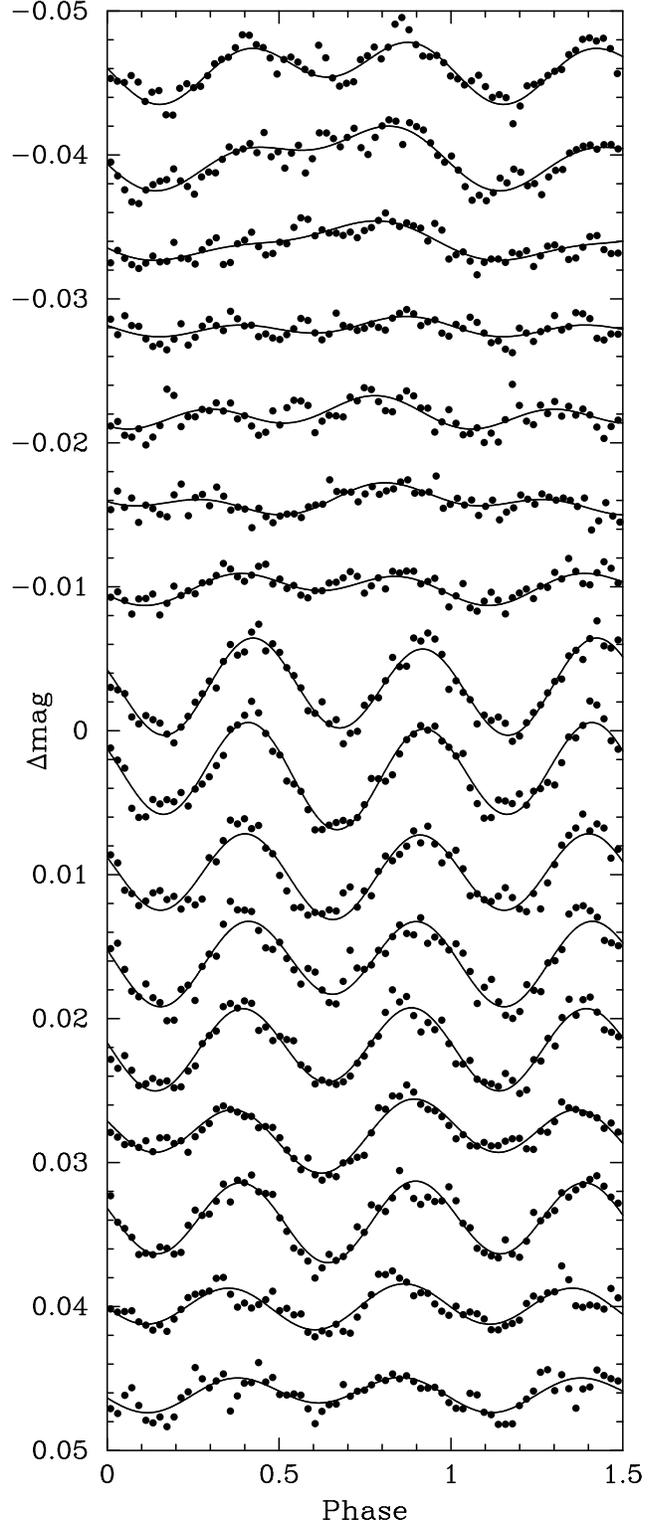}
\caption[]{Time series of phase diagrams, constructed from some three
years of LC {\it Kepler} photometry, folded around the subharmonic frequency. 
Each phase diagram comprises a $\sim 2$-month subset of data. Time runs 
from top to bottom. Fits to the phase plots are shown as well.}
\end{figure}

\subsection{Spectroscopic constraints}

The 1.23799-day {\it Kepler} photometric period is not detected in our 
spectroscopic data set. However, we caution that the temporal sampling of the 
spectra is not well suited to study a modulation on this time-scale. 
Nevertheless, phasing the optical spectra on 1.23799\,d does not provide 
a convincing result.

The stellar wind changes are extensive over short-time scales
(cf.\ Fig.\ 2). The He{\sc ii} 4686 emission equivalent width changes
by $\sim$ 30{\%} over 24 hours, and by $\sim$ 10{\%} in just
30 minutes. Even in the latter rapid fluctuation the outflow is
simultaneously altered over almost 300 km s$^{-1}$ blueward and redward 
of line centre, i.e. $\sim$ $\pm$ 0.25 of the wind terminal velocity in
NGC~6826. We suggest that the stochastic nature of the dynamic
spectra in Fig. 2 and the unstable frequencies evident in Fig. 4
point to the onset of clumping in the fast wind of the central star.

However, Fourier analysis of the optical spectroscopy suggests at least 
one potentially interesting modulation time scale, as demonstrated in 
Fig.\ 7. We determined the central velocities of the C{\sc IV} 5801\AA\, 
and 5812\AA\, emission lines by least squares Gaussian fits. The 
internal error bars on the velocities are $<1$ km/s, and are 
``normalised'' to the central velocity measurements of the $H_\beta$ 
nebular emission. The amplitude spectrum shows the strongest peak 
($f_{spec}$) at 7.821 \cd, or $P_{spec} \sim 3.1$\,hr, with $S/N = 
3.9$, just below our significance limit.

\begin{figure}
\includegraphics[width=81.6mm,viewport=00 33 270 715]{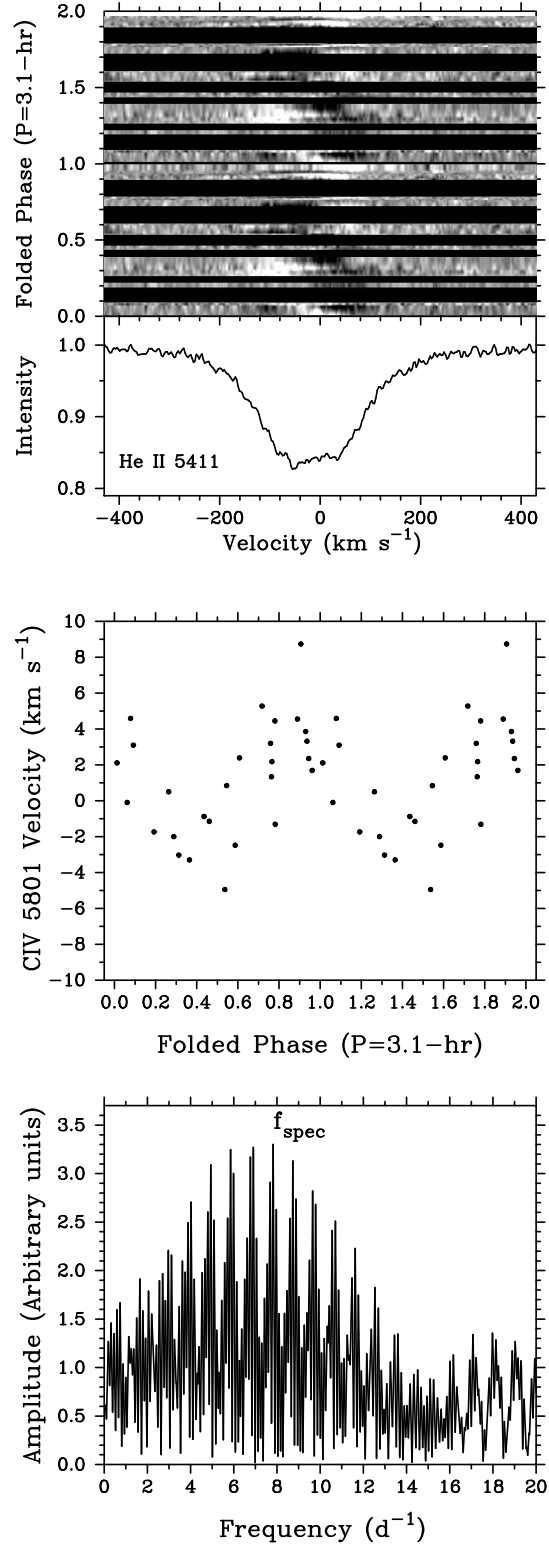}
\caption[]{Bottom panel: amplitude spectrum of radial velocity shifts
measured in the C {\sc IV} 5801\AA\, and 5812\AA\, emission lines.  
Middle panel: the C {\sc IV} central velocities phased on the 3.1-hr
period and folded over two cycles. Top panel, upper half: the combined
CFHT and NOT time series of the He {\sc II} 5411\AA\, absorption feature
shown as a dynamic spectrum now phased on the 3.1-hr period. Top
panel, lower half: the mean He {\sc II} 5411\AA\, line profile.}
\end{figure}

A phase diagram relative to this period (middle panel of Fig.\ 7) folds 
nicely as to be expected. In the uppermost panel of Fig.\ 7, the dynamic 
spectra of the He {\sc II} 5411\AA\, absorption feature have been folded 
on the 3.1-hr period. The phase coverage is reasonable, and the image 
suggests some evidence for organised behaviour, such as the red-to-blue 
enhancement of the flux with respect to the mean. Unfortunately, this 
3.1-hr spectral line modulation has no convincing counterpart in the 
{\it Kepler} photometry, even when only analysing the parts of the light 
curve coinciding with the spectroscopic observations (Fig.\ 8). Although 
we cannot rule out a binary system based solely on the optical 
spectroscopy, we nevertheless do not detect a consistent periodicity in 
the radial velocities of different emission and absorption lines in 
NGC~6826. The spectroscopic time series does not fold well with a 
0.619 or 1.238\,d period, which is at least partly due to the sampling 
of the data. For the same reason a 3.1-hr signal cannot be seen in Fig.\ 
2.

\begin{figure*}
\includegraphics[width=177mm,viewport=-80 3 600 215]{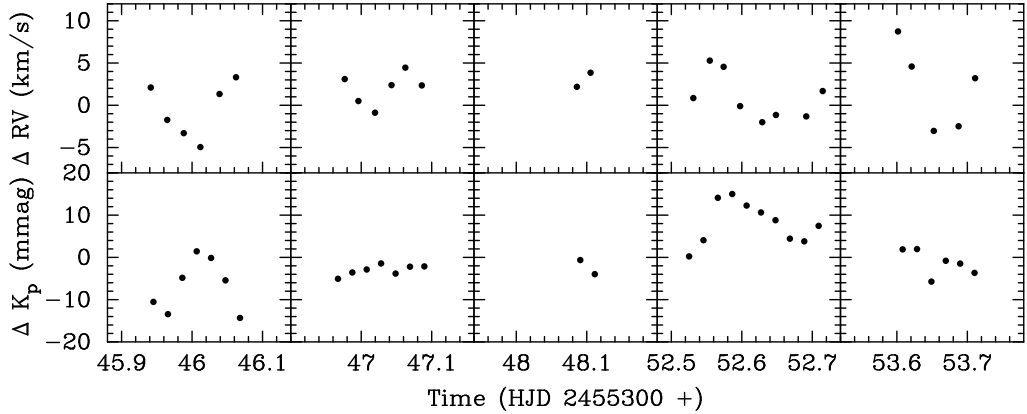}
\caption[]{Simultaneous C {\sc IV} radial velocities (upper panels) and 
{\it Kepler} photometry (lower panels) of NGC 6826. The non-overlapping parts 
of the light curves have been removed for easier comparison. The light 
and radial velocity variations do not correspond.}
\end{figure*}

\subsection{Summary of observational results}

The variability of NGC 6826 is characterized by:
\begin{itemize}
\item A basic $1.23799\pm0.00001$-d photometric modulation with a 
(mostly) double-wave variable light curve shape and total light range 
not exceeding 0.01~mag\footnote{As estimated in Sect.\ 2.1, the {\it 
intrinsic} light amplitude would be a factor of four or more higher 
due to the presence of the nebula.}.
\item Irregular variability with a light range of about 0.05~mag$^2$ and 
a time scale of a few hours.
\item Optical stellar absorption and emission lines variable on time 
scales of a few hours, a possible periodicity of 3.1 hr, with evidence 
for organized behaviour.
\item Discrete Absorption Components in wind-sensitive ultraviolet 
spectral lines, with at least two episodes occurring over 
$\sim$\,8.5\,hr (Prinja et al.\ 2012).
\item Red noise in the amplitude spectrum as a consequence of the 
irregular light variability.
\end{itemize}
There is no evidence for a spectroscopic counterpart of the photometric 
1.23799-d modulation, and no photometric evidence for the DAC 
re-occurrence time. Also, there is no correspondence between the 
optical radial velocity changes and the simultaneous {\it Kepler} 
photometry.

\section{Discussion}

The lack of radial velocity variations with the same period as the only
coherent signal in our data, its double-wave shape, and the temporal
changes of this shape strongly argue against a pulsational or binary
origin. The observed temporal behaviour of this signal is rather
consistent with rotational modulation, i.e.\ we see areas of different
surface brightness moving in and out of the line of sight. Prinja et al.\
(2012) measured $v \sin i = 50\pm10$~\kms for NGC 6826. With the
temperature ($T_{\rm eff}=46\,000$~K) and surface gravity
(log\,$g=3.8$) of the central star spectroscopically determined by
Kudritzki, Urbaneja \& Puls (2006), these authors obtained
$R=1.8$~R$_{\odot}$. This gives a maximum rotation period of about 1.8\,d.
The period of 1.23799\,d inferred above is consistent with this value, and
implies an inclination angle of the rotational axis of the order of
45\degr. This leaves sufficient horizon for brightness inhomogeneities to
appear and disappear. The shape of this photometric modulation and its
slow changes may be interpreted as two ``spots'' whose size or surface
brightness contrast somewhat vary with time.

Concerning the hourly photometric variability, the time scales 
themselves immediately rule out a binary or rotational origin; a 
potential secondary component would be located within the primary CSPN, 
or the CSPN would needs to rotate faster than breakup speed, 
respectively. The absence of any short-term periodicity in the whole 
data set and the irregularity of these variations argue that we do not 
see a photospheric phenomenon. This is also an argument that pulsations 
are not observed in the {\it Kepler} light curves: pulsation occurs at 
distinct frequencies. Even if it was stochastic, it can only appear and 
disappear at stellar eigenmode frequencies. Therefore one would expect 
to see at least some peaks standing out in the periodogram of the SC 
data (Figs.\ 3), which is not the case. However, we cannot claim that 
pulsations are not present at all because the photometry contains little 
flux that originates in the stellar photosphere.

We evaluate the time scales of possible central star pulsations. From 
the most suitable sequence of envelope models by Gautschy (1993) with 
respect to NGC 6826 we estimate pulsation constant 
$Q=P\sqrt{\rho/\rho_{\sun}}=0.042$\,d for the fundamental radial mode, 
that is excited by the $\kappa$-mechanism in almost all of these models. 
This pulsation constant corresponds to a period of the radial 
fundamental mode of NGC 6826 of 2.7~hr. Given the uncertainties in the 
stellar parameters and models, this can only be seen as a rough 
estimate. We note that envelope models do not allow the computation of 
nonradial modes, but the presence of strange modes with similar periods 
as the radial pulsations was noticed in the models by Gautschy (1993).

It is important to constrain under which circumstances we would 
have missed the detection of binary-induced variability. A binary would 
reveal itself through radial velocity or light variability. The 
photometric noise level in our data is 0.072 mmag at worst. This 
translates into a detection threshold for periodic signals of 0.29 mmag. 
Given the dilution of any given variability signal due to the presence 
of the nebula (Sect.\ 2.1), we conservatively estimate that we would 
have detected any periodic signal with an intrinsic amplitude exceeding 
1.5 mmag in our data.

Morris (1985) derived expressions for the amplitude of ellipsodial light 
variations. We applied these in combination with the limb darkening and 
gravity brightening coefficients for the Kepler bandpass by Claret \& 
Bloemen (2011), and adopting a central star mass of 0.74 M$_{\odot}$ 
(Kudritzki et al.\ 2006). This resulted in the constraint that we could 
have detected binaries with orbital periods shorter than seven days 
under favourable inclinations. However, a stellar 
($M\geq0.08$\,M$_{\odot}$) companion in a 1.238-d orbit would only have 
escaped our detection in combination with an orbital inclination below 
26\degr.

Under the assumption of a random orientation of the orbital plane 
in space, the probability that it is observed below a certain
inclination angle is
\begin{equation}
p(<i)=1-\cos i.
\end{equation}
Further assuming that the CSPN rotation axis is normal to this orbital 
place, this probability can be written as
\begin{equation}
\begin{array}{lr}
p(<i)=\frac{v_b(1-\cos i)}{\sqrt{v_b^2-(v\sin i)^2}} & 
i\geq\sin^{-1}((v\sin i)/v_b)\\
p(<i)=0 & 0\leq i<\sin^{-1}((v\sin i)/v_b),
\end{array}
\end{equation}
where $v_b$ is the stellar rotational break-up velocity (see Bernacca 
(1970) for a detailed discussion).

We can estimate the breakup velocity of the CSPN by using the 
spectroscopic parameters quoted earlier and
\begin{equation}
P_{crit} = \frac{2\pi R_{eq}^{3/2}}{(GM)^{1/2}},
\end{equation}
where $R_{eq}$ is the equatorial radius of the star. Following Reid 
et al.\ (1993), $R_{eq} = 1.5\,R$, where $R$ is the stellar radius 
in the non-rotating case. This calculation yields $P_{crit}=7.8$\,h, 
hence $v_b=280$\,\kms, hence $\sin^{-1}((v\sin i)/v_b)=10.3$\degr. With 
these values, probabilities calculated with Eq.\ 2 are only 1.6\% higher 
as if they were derived with Eq.\ 1.
Returning to the hypothesized stellar companion in an orbit inclined 
by less than 26\degr discussed earlier, the probability that we missed 
its detection therefore becomes 10\%.

A stronger constraint arises from the lack of a detection of 
periodic radial velocity changes. We again used a central star mass of 
0.74 M$_{\odot}$, and a generous upper limit for its periodic radial 
velocity variations of 3.3 \kms. For orbital periods shorter than seven 
days (the length of our spectroscopic time series), we should have 
detected all stellar companions with an orbital inclination above 
44\degr; for hypothesized companions exceeding 0.2\,M$_{\odot}$, the 
limit on the orbital inclination is $i<31$\degr, corresponding to a 14\% 
probability of missing a detection. A stellar companion in a 1.238-d 
orbit would not have been detected in radial velocity with an orbital 
inclination below 10\degr. However, in that case the CSPN 
rotation axis cannot be normal to the orbital plane because the central 
star would have to rotate above break-up speed.

The referee remarked that the shape of the NGC 6826 nebula and the 
presence of Fast-Low-Ionization Emission Regions (FLIERs) in it (Balick 
et al.\ 1994) suggest that the orbital plane of a hypothetical binary 
causing these features would be at low inclination. Whereas such an 
interpretation cannot be ruled out completely, it needs to be reconciled 
with the orientation of the CSPN's rotation axis keeping in mind its 
critical rotation rate. Furthermore, such a low inclination is 
inconsistent with our interpretation of the 1.238-d variation in terms 
of a rotational modulation originating from the central star, unless one 
postulates a rotation axis significantly deviating from the normal to
the orbital plane.

\section{Interpretation}

Following their analysis of one year of {\it Kepler} LC data on NGC 
6826, Jevti\'{c} et al.\ (2012) suggested that a combination of stellar 
pulsation and interaction with a close companion is a possible 
explanation of the central star's variability. According to our results, 
such a scenario is very unlikely. Our measurements rather support 
another possibility mentioned by Jevti\'{c} et al.\ (2012), variable 
features on the central star's surface in combination with 
inhomogeneities and variations in the density structure of its wind.
Such variability is observationally well established in massive hot 
stars (e.g., Fullerton 2003).

Large-scale variations in the wind structure are believed to be seated 
at the base of the wind, in so-called Co-Rotating Interaction Regions 
(CIRs, Mullan 1984). In brief, CIRs arise due to intensity variations on 
or near the stellar photosphere, such as spots or nonradial pulsation 
patterns. Such intensity or temperature variations modify the outflow 
velocity of the spherical wind. Hence, the parts of the wind originating 
at these locations are differently affected by the driving force, and 
collide with particles emitted from other surface regions. Due to 
stellar rotation spiral-shaped density fluctuations in the wind arise 
and cause the spectroscopically observed DACs. These DACs occur on time 
scales related to the stellar rotation period (several days), and drift 
slowly through the line profiles with respect to the wind velocity. As 
exemplary literature we refer to Fullerton et al.\ (1997) for an 
observational study and Cranmer \& Owocki (1996) and Lobel \& Blomme 
(2008) for hydrodynamical model computations.

In addition to the DACs, ``modulations'' have also been observed to move 
through the line profiles. They travel at a much faster rate than the 
DACs and have considerably shorter recurrence times (Fullerton et al.\ 
1997). These ``modulations'' alter the mean flux, as opposed to the DACs 
that are pure absorption features. Radiative transfer modelling (Lobel, 
Toal\'a \& Blomme 2011) implies that these are only slightly bowed
large-scale density enhancements and local wind velocity variations that 
radially protrude into the equatorial wind. The density enhancement is 
of the order of 10\%, and can result from mechanical wave action at the 
base of the stellar wind, such as produced by nonradial pulsations 
(e.g., see Kaufer et al.\ 2006). The photospheric cause of these 
variations is called Rotational Modulation Regions (RMRs).

The variability of some Wolf-Rayet (WR) stars may be related to such 
variations, although they have been interpreted slightly differently. 
Periodic photometric variations with time scales of days have been 
attributed to CIRs (e.g., Chen\'e et al.\ 2011), and irregular 
short-term variations with time scales of hours have been reported in 
addition. Their spectroscopic counterparts are believed to be moving 
features in optical profiles of wind-sensitive lines, observed not only 
in WR stars, but also in Of stars (e.g., L\'epine \& Moffat 2008) and 
are interpreted as clumps propagating in the wind.

The temporal behaviour of NGC 6826 conforms to these scenarios. Taking 
the photometric 1.23799\,d period to be due to rotation, the double-wave 
shape may be a manifestation of two co-rotating features. The 
recurrence time for DACs in the UV lines of NGC~6826 is not well 
constrained; Prinja et al.\ (2012) reported the detection of two 
sequential features over $\sim$ 8 hours. It is interesting to note 
however that we never see more than two strong CIRs in OB stars that 
have been monitored for one rotation period.

%The photometric period could 
%be explained in terms of a CIR moving prograde with the rotational 
%motion. For the massive O star HD 64760 (Lobel \& Blomme 2008) a 
%counter-rotating CIR was proposed. Such an interpretation would 
%immediately rule out a magnetic field as the cause for the CIR, as it 
%would be locked to the stellar rotation period. It however remains to 
%be shown that there is a causal connection between the 1.238-d 
%photometric period and CIRs.

The time scales of short term light variations are consistent with RMRs 
or clumps moving within the wind. As our spectroscopy implies that the 
variations in the wind are not local, they are most likely attributable 
to large-scale features. Therefore they may be associated with RMRs. The 
difficulty with this picture is that the spectroscopic data of massive 
stars, in which RMRs are manifested, imply that they occur quite 
regularly, whereas the short-term photometric variability is irregular 
in our case. Consequently, some periodically triggered mass loss must 
have lost memory by the time it reaches the depth in the wind that the 
photometry samples.

Albeit the present observations do give some insight into the nature 
of the variability of NGC 6826 (and other ZZ Lep stars), more 
observational evidence is needed. Although observationally hard to 
obtain due to the faintness of the targets, spectroscopic evidence for 
wind clumping or eventual RMR presence in CSPN should be sought. On the 
other hand, photometric evidence for light variability on the RMR and 
CIR time scales for massive stars, focusing on non-WR stars, would also 
be useful.

In this context it is interesting to note that Blomme et al.\ (2011) 
analysed CoRoT light curves of three massive O-type stars and also found 
some apparently incoherent variability, in all three targets. The light 
curves, amplitude spectra, and time-frequency analyses of these stars 
phenomenologically closely resemble what we report for NGC~6826. One may 
therefore speculate that this kind of variability is present in all hot 
stars (Blomme et al.\ (2011) and Handler et al.\ (2012) discussed it in 
connection with massive OB stars), and that it is related to stellar 
wind variations.

\section{Summary}

The observed photometric variability of the central star of NGC 6826
consists of a periodic modulation with a time scale of 1.23799\,d and
irregular light variations on time scales of a few hours. Optical spectra 
imply line variability on a similar time scale, with a possible period of 
3.1 hr. Ultraviolet spectra show DACs with a re-occurrence time of about 
8.5 hr. The only possible correspondence between the photometric and 
spectroscopic variability may be the time scale of the hourly changes.

The periodic photometric variation is best explained by rotational 
modulation. The 1.23799\,d-period we derived is similar to asteroseismic 
rotation periods of single white dwarf stars (e.g., Winget et al.\ 
1991). Time-series photometry of ZZ Lep stars may therefore reveal their 
rotation periods, but large data sets are needed to detect these 
periodicities due to the dominating short-term variability, that is due 
to changes in the stellar mass loss. This behaviour is similar to what 
has already been observed in massive OB and Wolf-Rayet stars, suggesting 
that the same mechanism may be responsible for the variations in all 
hot-star winds.

\section*{ACKNOWLEDGEMENTS}

Funding for this Discovery mission is provided by NASA's Science Mission 
Directorate. The authors thank the {\it Kepler} team for their 
continuous work that ensures the best possible science output. Funding 
for the Stellar Astrophysics Centre (SAC) is provided by The Danish 
National Research Foundation. The research is supported by the ASTERISK 
project (ASTERoseismic Investigations with SONG and {\it Kepler}) funded 
by the European Research Council (Grant agreement no.: 267864). The 
authors would like to thank S.\ Simon-Diaz for his assistance with the 
FIES/NOT observations. GH acknowledges funding through NCN grant 
2011/01/B/ST9/05448, and thanks Nada Jevti\'{c} and Andrzej Baran for 
their comments on a draft version of this paper. Comments by the 
anonymous referee improved part of our argumentation. This article is 
partly based on observations made with the Nordic Optical Telescope 
operated on the island of La Palma by the Nordic Optical Telescope 
Scientific Association in the Spanish Observatorio del Roque de los 
Muchachos.

\bsp

\end{document}